\documentclass[letterpaper, 10 pt, conference]{ieeeconf}  %

\IEEEoverridecommandlockouts                              %
\usepackage{amsmath} %
\usepackage{amssymb}  %

\usepackage{graphicx}
\usepackage[acronym]{glossaries}
\usepackage{bm}
\usepackage{multirow}
\usepackage{xfrac}
\usepackage{tikz}
\usetikzlibrary{positioning}
\usetikzlibrary{fit,backgrounds,bending}
\usetikzlibrary{external}
\usepackage{tabularx}
\usepackage{arydshln}
\usepackage{siunitx}
\let\labelindent\relax
\usepackage[inline]{enumitem}
\usepackage{pgf}
\usepackage{pgfplots}
\pgfplotsset{compat=1.11}
\usepgfplotslibrary{units}
\usepackage{nicematrix}
\usepackage{calc}
\usepackage{paralist}
\usepackage{enumitem}
\makeatletter
\let\NAT@parse\undefined
\makeatother
\usepackage[hidelinks]{hyperref}
\usepackage{etoolbox}

\usepackage{xcolor}

\newcommand{\videourl}{\url{https://youtu.be/lrrWBx2ASaE}}%

\newcommand*{\R}{\mathbb{R}}
\newcommand*{\N}{\mathbb{N}}

\newcommand*{\genIdx}{\genIdxChar}
\newcommand*{\genDim}{\genDimChar}

\newcommand*{\genIdxChar}{j}
\newcommand*{\genDimChar}{r}
\newcommand*{\genDelChar}{\tau}
\newcommand*{\genConst}{\kappa}

\newcommand*{\timeChar}{t}
\newcommand*{\timestepChar}{k}
\newcommand*{\stateChar}{x}
\newcommand*{\stateDimChar}{S}
\newcommand*{\inpChar}{u}
\newcommand*{\outpChar}{y}
\newcommand*{\distChar}{d}
\newcommand*{\sysEqChar}{\alpha}
\newcommand*{\outpEqChar}{\beta}
\newcommand*{\sysMatChar}{A}
\newcommand*{\inpMatChar}{B}
\newcommand*{\distMatChar}{D}
\newcommand*{\offVecChar}{f}
\newcommand*{\outpVecChar}{c}%

\newcommand*{\llmIdxChar}{i}
\newcommand*{\llmNumChar}{K}
\newcommand*{\weightsChar}{\theta}
\newcommand*{\inpNumChar}{P}%
\newcommand*{\actChar}{\Phi}
\newcommand*{\valChar}{\phi}

\newcommand*{\inpDelChar}{m}
\newcommand*{\outpDelChar}{n}
\newcommand*{\distDelChar}{q}
\newcommand*{\inpDelNumChar}{M}
\newcommand*{\outpDelNumChar}{N}
\newcommand*{\distDelNumChar}{Q}
\newcommand*{\inpDelsChar}{\mathcal{M}}
\newcommand*{\outpDelsChar}{\mathcal{N}}
\newcommand*{\distDelsChar}{\mathcal{Q}}
\newcommand*{\lmnFunChar}{\Psi}

\newcommand*{\relDegChar}{\delta}

\newcommand*{\transInpChar}{v}

\newcommand*{\inpParamChar}{b}
\newcommand*{\outpParamChar}{a}
\newcommand*{\offParamChar}{c}
\newcommand*{\identityChar}{I}
\newcommand*{\desChar}{w}

\newcommand*{\tcpChar}{\chi}
\newcommand*{\cylPosChar}{s}

\newcommand*{\discrChar}{z}
\newcommand*{\poleChar}{z}%
\newcommand*{\inpZChar}{U}
\newcommand*{\transInpZChar}{V}

\newcommand*{\facMatChar}{\Gamma}
\newcommand*{\facChar}{\gamma}
\newcommand*{\posDefMatChar}{P}
\newcommand*{\shiftChar}{\Delta}

\newcommand*{\dInpIdxChar}{g}
\newcommand*{\dOutpIdxChar}{h}
\newcommand*{\dInpNumChar}{G}
\newcommand*{\dOutpNumChar}{H}
\newcommand*{\outpMatChar}{C}

\newcommand*{\sysTime}{\timeChar}
\newcommand*{\sysSubsrc}{sys}
\newcommand*{\sysState}[1][\sysTime]{\bm{\stateChar}_{\text{\sysSubsrc}}(#1)}
\newcommand*{\sysStateChange}[1][\sysTime]{\bm{\dot{\stateChar}}_{\text{\sysSubsrc}}(#1)}
\newcommand*{\sysInp}[1][\sysTime]{\bm{\inpChar}_{\text{\sysSubsrc}}(#1)}
\newcommand*{\sysOutp}[1][\sysTime]{\bm{\outpChar}_{\text{\sysSubsrc}}(#1)}
\newcommand*{\sysDist}[1][\sysTime]{\distChar_{\text{\sysSubsrc}}(#1)}
\newcommand*{\sysSysEq}[1]{\sysEqChar_{\text{\sysSubsrc}}(#1)}
\newcommand*{\sysOutpEq}[1]{\outpEqChar_{\text{\sysSubsrc}}(#1)}

\newcommand*{\valSubscr}{nl}
\newcommand*{\linSubscr}{lin}
\newcommand*{\lmnOutp}[1][\lmnLinInpVec,\lmnActInpVec]{{\lmnFunChar}(#1)}
\newcommand*{\lmnNumModels}{\llmNumChar}
\newcommand*{\lmnModelIdx}{\llmIdxChar}
\newcommand*{\lmnModelOutp}[2][\lmnLinInpVec]{{\lmnFunChar}_{#2}(#1)}
\newcommand*{\lmnModelWeight}[2]{\weightsChar_{#1#2}}

\newcommand*{\lmnModelAct}[2][\lmnActInpVec]{\actChar_{#2}(#1)}
\newcommand*{\lmnActVec}[1][\lmnActInpVec]{\bm{\actChar}(#1)}
\newcommand*{\lmnActInpVec}{\bm{\inpChar}_{\text{\valSubscr}}}
\newcommand*{\lmnLinInpVec}{\bm{\inpChar}_{\text{\linSubscr}}}

\newcommand*{\lmnLinInp}[1]{\inpChar_{\text{\linSubscr},#1}}
\newcommand*{\lmnLinInpSize}{\inpNumChar}
\newcommand*{\lmnActInpSize}{\Tilde{\inpNumChar}}

\newcommand*{\lmnModelVal}[1]{\valChar_{#1}(\lmnActInpVec)}

\newcommand*{\timestep}{\timestepChar}
\newcommand*{\narxOutp}[1][\timestep]{\hat{\outpChar}(#1)}
\newcommand*{\narxInp}[1][\timestep]{\inpChar(#1)}
\newcommand*{\narxDist}[1][\timestep]{\distChar(#1)}
\newcommand*{\narxLmn}[1][\narxLinInpVec,\narxActInpVec]{\lmnOutp[#1]}
\newcommand*{\narxInpLinDel}[1]{\inpDelChar_{#1}}
\newcommand*{\narxOutpLinDel}[1]{\outpDelChar_{#1}}
\newcommand*{\narxDistLinDel}[1]{\distDelChar_{#1}}
\newcommand*{\narxInpActDel}[1]{\Tilde{\inpDelChar}_{#1}}
\newcommand*{\narxOutpActDel}[1]{\Tilde{\outpDelChar}_{#1}}
\newcommand*{\narxDistActDel}[1]{\Tilde{\distDelChar}_{#1}}
\newcommand*{\narxInpLinDelSet}{\inpDelsChar}
\newcommand*{\narxOutpLinDelSet}{\outpDelsChar}
\newcommand*{\narxDistLinDelSet}{\distDelsChar}
\newcommand*{\narxInpActDelSet}{\Tilde{\inpDelsChar}}
\newcommand*{\narxOutpActDelSet}{\Tilde{\outpDelsChar}}
\newcommand*{\narxDistActDelSet}{\Tilde{\distDelsChar}}
\newcommand*{\narxLinInpVec}[1][\timestep]{\lmnLinInpVec(#1)}
\newcommand*{\narxActInpVec}[1][\timestep]{\lmnActInpVec(#1)}
\newcommand*{\narxMaxInpDel}{\inpDelNumChar}
\newcommand*{\narxMaxOutpDel}{\outpDelNumChar}
\newcommand*{\narxMaxDistDel}{\distDelNumChar}

\newcommand*{\relDeg}{\relDegChar}

\newcommand*{\dlmnStateVec}[1][\timestep]{\bm{\stateChar}(#1)}
\newcommand*{\dlmnInp}[1][\timestep]{\inpChar(#1)}
\newcommand*{\dlmnOutp}[1][\timestep]{\hat{\outpChar}(#1)}
\newcommand*{\dlmnNumStates}{\stateDimChar}
\newcommand*{\dlmnActVec}[1][\timestep]{\bm{\actChar}_{#1}}
\newcommand*{\dlmnAct}[1]{\actChar_{#1}(\timestep)}
\newcommand*{\dlmnCombSysMat}[1][\timestep]{\bm{\sysMatChar}(\dlmnActVec[#1])}
\newcommand*{\dlmnModelSysMat}[1]{\bm{\sysMatChar}_{#1}}
\newcommand*{\dlmnCombInpMat}[1][\timestep]{\bm{\inpMatChar}(\dlmnActVec[#1])}
\newcommand*{\dlmnModelInpMat}[1]{\bm{\inpMatChar}_{#1}}
\newcommand*{\dlmnCombOffVec}[1][\timestep]{\bm{\offVecChar}(\dlmnActVec[#1])}
\newcommand*{\dlmnModelOffVec}[1]{\bm{\offVecChar}_{#1}}
\newcommand*{\dlmnOutpMat}{\bm{\outpVecChar}^\top}

\newcommand*{\dlmnInpDel}{\narxInpLinDel{}}
\newcommand*{\dlmnOutpDel}{\narxOutpLinDel{}}

\newcommand*{\dlmnCombInpParam}[1]{\inpParamChar_{#1}(\dlmnActVec)}
\newcommand*{\dlmnCombOutpParam}[1]{\outpParamChar_{#1}(\dlmnActVec)}
\newcommand*{\dlmnCombOffParam}{\offParamChar(\dlmnActVec)}

\newcommand*{\dlmnCombInpParamTVec}{\bm{\inpParamChar}^\top(\dlmnActVec)}
\newcommand*{\dlmnCombOutpParamTVec}{\bm{\outpParamChar}^\top(\dlmnActVec)}
\newcommand*{\dlmnModelInpParamTVec}[1]{\bm{\inpParamChar}^{(#1)\top}}
\newcommand*{\dlmnModelOutpParamTVec}[1]{\bm{\outpParamChar}^{(#1)\top}}
\newcommand*{\dlmnModelInpParam}[2]{\inpParamChar_{#1}^{(#2)}}
\newcommand*{\dlmnModelOutpParam}[2]{\outpParamChar_{#1}^{(#2)}}
\newcommand*{\dlmnModelOffParam}[1]{\offParamChar^{(#1)}}
\newcommand*{\dlmnModelInpParamNoMod}[1]{\inpParamChar_{#1}}
\newcommand*{\dlmnModelOutpParamNoMod}[1]{\outpParamChar_{#1}}
\newcommand*{\dlmnModelOffParamNoMod}{\offParamChar}

\newcommand*{\upperShiftMatrix}[1]{\bm{\underbar{\identityChar}}_{#1}}
\newcommand*{\identitySub}[1]{\bm{\Bar{\identityChar}}_{#1}}

\newcommand*{\dlmnDes}[1][\timestep]{\desChar(#1)}
\newcommand*{\dlmnStateVecDes}[1][\timestep]{\bm{\stateChar}_{\desChar}(#1)}
\newcommand*{\dlmnDesShift}[1][\timestep]{\transInpChar(#1)}
\newcommand*{\dlmnStateVecDesShift}[1][\timestep]{\bm{\stateChar}_{\transInpChar}(#1)}

\newcommand*{\genDel}{\genDelChar}

\newcommand*{\excMeasTcp}{\bm{\tcpChar}_{\text{meas}}}
\newcommand*{\excDesTcp}{\bm{\tcpChar}_{\text{des}}}
\newcommand*{\excDesTcpVel}{\bm{\dot{\tcpChar}}_{\text{des}}}
\newcommand*{\excMeasCyl}{\bm{\cylPosChar}_{\text{meas}}}

\newcommand*{\excDesCylVel}{\bm{\dot{\cylPosChar}}_{\text{des}}}
\newcommand*{\excPress}{\bm{\distChar}_{\text{press}}}
\newcommand*{\excJoyPos}{\bm{\inpChar}_{\text{joy}}}

\newcommand*{\discrVar}{\discrChar}
\newcommand*{\pole}[1]{\poleChar_{#1}}
\newcommand*{\dlmnInpZ}{\inpZChar(\discrVar)}
\newcommand*{\dlmnDesShiftZ}{\transInpZChar(\discrVar)}

\newcommand*{\invLmnStateVec}[1][\timestep]{\bm{\hat{\stateChar}}(#1)}
\newcommand*{\invLmnStateVecEq}{\bm{\hat{\stateChar}}_{0}}
\newcommand*{\invLmnCombSysMat}[1][\dlmnActVec]{\bm{\hat{\sysMatChar}}(#1)}
\newcommand*{\invLmnModelSysMat}[1]{\hat{\bm{\sysMatChar}}_{#1}}
\newcommand*{\invLmnCombInpMat}[1][\dlmnActVec]{\bm{\hat{\inpMatChar}}(#1)}
\newcommand*{\invLmnModelInpMat}[1]{\hat{\bm{\inpMatChar}}_{#1}}
\newcommand*{\invLmnCombOffVec}[1][\dlmnActVec]{\bm{\hat{\offVecChar}}(#1)}
\newcommand*{\invLmnModelOffVec}[1]{\bm{\hat{\offVecChar}}_{#1}}
\newcommand*{\invLmnOutpMat}{\bm{\hat{\outpVecChar}}^\top}
\newcommand*{\invLmnDesShiftVec}[1][\timestep]{\bm{\hat{\transInpChar}}(#1)}
\newcommand*{\invLmnNumStates}{\stateDimChar}
\newcommand*{\invLmnCombFacMat}{\bm{\facMatChar}(\dlmnActVec)}
\newcommand*{\invLmnCombFac}{\facChar(\dlmnActVec)}
\newcommand*{\invLmnModelFacMat}[1]{\bm{\facMatChar}_{#1}}
\newcommand*{\invLmnModelFac}[1]{\facChar_{#1}}

\newcommand*{\posDefMat}{\bm{\posDefMatChar}}

\newcommand*{\distSubscr}{d}
\newcommand*{\dlmnStateVecDist}[1][\timestep]{\bm{\stateChar}_{\text{\distSubscr}}(#1)}
\newcommand*{\dlmnDist}[1][\timestep]{\distChar(#1)}
\newcommand*{\dlmnCombSysMatDist}[1][\timestep]{\bm{\sysMatChar}_{\text{\distSubscr}}(\dlmnActVec[#1])}
\newcommand*{\dlmnModelSysMatDist}[1]{\bm{\sysMatChar}_{\text{\distSubscr},#1}}
\newcommand*{\dlmnCombInpMatDist}[1][\timestep]{\bm{\inpMatChar}_{\text{\distSubscr}}(\dlmnActVec[#1])}
\newcommand*{\dlmnModelInpMatDist}[1]{\bm{\inpMatChar}_{\text{\distSubscr},#1}}
\newcommand*{\dlmnCombDistMatDist}[1][\timestep]{\bm{\distMatChar}_{\text{\distSubscr}}(\dlmnActVec[#1])}
\newcommand*{\dlmnModelDistMatDist}[1]{\bm{\distMatChar}_{\text{\distSubscr},#1}}
\newcommand*{\dlmnCombOffVecDist}[1][\timestep]{\bm{\offVecChar}_{\text{\distSubscr}}(\dlmnActVec[#1])}
\newcommand*{\dlmnModelOffVecDist}[1]{\bm{\offVecChar}_{\text{\distSubscr},#1}}
\newcommand*{\dlmnNumStatesDist}{\stateDimChar}
\newcommand*{\dlmnStateVecDesDist}[1][\timestep]{\bm{\stateChar}_{\text{\distSubscr\desChar}}(#1)}
\newcommand*{\dlmnDistDel}{\narxDistLinDel{}}
\newcommand*{\relDegDist}{\relDegChar_{\text{\distSubscr}}}

\newcommand*{\dlmnModelDistParamTVec}[1]{\bm{\distChar}^{(#1)\top}}
\newcommand*{\dlmnModelDistParam}[2]{\distChar_{#1}^{(#2)}}

\newcommand*{\mimoSubscr}{m}
\newcommand*{\dlmnInpIdx}{\dInpIdxChar}
\newcommand*{\dlmnOutpIdx}{\dOutpIdxChar}
\newcommand*{\dlmnMaxInp}{\dInpNumChar}
\newcommand*{\dlmnMaxOutp}{\dOutpNumChar}

\newcommand*{\dlmnActVecMimoComb}[1][\timestep]{\bm{\actChar}_{\text{\mimoSubscr},#1}}
\newcommand*{\dlmnInpMimo}[2][\timestep]{\inpChar_{#2}(#1)}
\newcommand*{\dlmnOutpMimo}[2][\timestep]{\hat{\outpChar}_{#2}(#1)}
\newcommand*{\dlmnInpVecMimo}[1][\timestep]{\bm{\inpChar}(#1)}
\newcommand*{\dlmnOutpVecMimo}[1][\timestep]{\bm{\hat{\outpChar}}(#1)}
\newcommand*{\dlmnDesVecMimo}[1][\timestep]{\bm{\desChar}(#1)}
\newcommand*{\dlmnMaxInpDelMimo}[1]{\inpDelNumChar_{#1}}
\newcommand*{\dlmnMaxOutpDelMimo}[1]{\outpDelNumChar_{#1}}
\newcommand*{\dlmnStateVecMimo}[1][\timestep]{\bm{\stateChar}_\text{\mimoSubscr}(#1)}
\newcommand*{\dlmnCombSysMatMimo}[1][\timestep]{\bm{\sysMatChar}_\text{\mimoSubscr}(\dlmnActVecMimoComb[#1])}
\newcommand*{\dlmnCombInpMatMimo}[1][\timestep]{\bm{\inpMatChar}_\text{\mimoSubscr}(\dlmnActVecMimoComb[#1])}
\newcommand*{\dlmnCombOffVecMimo}[1][\timestep]{\bm{\offVecChar}_\text{\mimoSubscr}(\dlmnActVecMimoComb[#1])}
\newcommand*{\dlmnOutpMatMimo}{\bm{\outpMatChar}_\text{\mimoSubscr}}
\newcommand*{\dlmnNumStatesMimo}{\stateDimChar}

\newcommand*{\vtActVecA}{\bm{\actChar}}
\newcommand*{\vtActVecB}{\bm{\bar{\actChar}}}
\newcommand*{\vtActA}[1]{\actChar_{#1}}
\newcommand*{\vtActB}[1]{\bar{\actChar}_{#1}}
\newcommand*{\vtParam}[2]{\inpParamChar_{#1}^{(#2)}}
\newcommand*{\vtParamVec}[1]{\bm{\inpParamChar}^{(#1)}}

\newcommand*{\tsfmStateVec}[1][\timestep]{\bm{\bar{\stateChar}}(#1)}
\newcommand*{\tsfmModelSysMat}[1]{\bm{\bar{\sysMatChar}}_{#1}}
\newcommand*{\tsfmAct}[1]{\actChar_{#1}}
\newcommand*{\tsfmNumStates}{\stateDimChar}

\newcommand*{\invLmnActAlt}[1]{\bar{\actChar}_{#1}(\timestep)}
\newcommand*{\invLmnActAltVec}[1][\timestep]{\bm{\bar{\actChar}}_{#1}}

\newcommand*{\invLmnCombInp}{\bm{\hat{f}}_\text{s}(\dlmnActVec, \invLmnDesShiftVec)}
\newcommand*{\invLmnStateVecShifted}[1][\timestep]{\bm{\hat{\stateChar}}_\text{s}(#1)}
\newcommand*{\invLmnStateVecShift}{\bm{\shiftChar\hat{\stateChar}}(\timestep)}

\newacronym{lmn}{LMN}{local model network}
\newacronym{narx}{NARX}{nonlinear autoregressive with exogenous input}
\newacronym{siso}{SISO}{single-input single-output}
\newacronym{miso}{MISO}{multiple-input single-output}
\newacronym{mimo}{MIMO}{multiple-input multiple-output}
\newacronym{lpv}{LPV}{linear parameter-varying}
\newacronym{tcp}{TCP}{Tool Center Point}
\newacronym{mlp}{MLP}{multilayer perceptron}
\newacronym{gpr}{GP-R}{Gaussian process regression}
\newacronym{llm}{LLM}{local linear model}
\newacronym{lolimot}{LOLIMOT}{local linear model tree}
\newacronym{tsfm}{TSFM}{Takagi-Sugeno fuzzy models}
\newacronym{nfir}{NFIR}{nonlinear finite impulse response}
\newacronym{sgd}{SGD}{stochastic gradient descent}
\newacronym{bibo}{BIBO}{bounded-input, bounded-output}
\newacronym{gas}{GAS}{globally asymptotically stable}
\newacronym{iss}{ISS}{input-to-state stability}
\newacronym{zd}{ZD}{zero dynamics}

\newtheorem{theorem}{Theorem}
\newtheorem{lemma}{Lemma}
\newtheorem{remark}{Remark}
\newtheorem{assumption}{Assumption}
\newtheorem{definition}{Definition}

\newlength{\tabcoldots}
\newlength{\arraycolsepdots}
\usepackage{fancyhdr}
\newcommand{\mytitle}{To appear in the proceedings of the
\textit{63rd IEEE Conference on Decision and Control}.\\
\copyright 2024 IEEE. Personal use of this material is permitted. Permission from IEEE must be obtained for all other uses, in any current or future media, including reprinting/republishing this material for advertising or promotional purposes, creating new collective works, for resale or redistribution to servers or lists, or reuse of any copyrighted component of this work in other works.}
\title{\LARGE \bf
Feedforward Controllers from Learned Dynamic Local Model Networks with Application to Excavator Assistance Functions 
}

\author{Leon Greiser, Ozan Demir, Benjamin Hartmann, Henrik Hose, and Sebastian Trimpe%
\thanks{This work was supported by the Robert Bosch GmbH, Stuttgart, Germany and in part by the Deutsche Forschungsgemeinschaft (DFG, German Research Foundation) – 2236/2 RTG "UnRAVeL".}%
\thanks{L.~Greiser, O.~Demir, and B.~Hartmann are with the Robert Bosch GmbH, Stuttgart, Germany {\tt\small \{ozan.demir,benjamin.hartmann\}@de.bosch.com}}%
\thanks{L.~Greiser, H.~Hose, and S.~Trimpe are with the Institute for Data Science in Mechanical Engineering, RWTH Aachen University, Germany\hspace{20em} {\tt\small leon.greiser@rwth-aachen.de} {\tt\small \{henrik.hose,trimpe\}@dsme.rwth-aachen.de}}%
}

\begin{document}

\maketitle
\thispagestyle{fancy}
\pagestyle{fancy}

\begin{abstract}
Complicated first principles modelling and controller synthesis can be prohibitively slow and expensive for high-mix, low-volume products such as hydraulic excavators. 
Instead, in a data-driven approach, recorded trajectories from the real system can be used to train local model networks (LMNs), for which feedforward controllers are derived via feedback linearization.
However, previous works required LMNs without zero dynamics for feedback linearization, which restricts the model structure and thus modelling capacity of LMNs.
In this paper, we overcome this restriction by providing a criterion for when feedback linearization of LMNs with zero dynamics yields a valid controller. As a criterion we propose the bounded-input bounded-output stability of the resulting controller.
In two additional contributions, we extend this approach to consider measured disturbance signals and multiple inputs and outputs.
We illustrate the effectiveness of our contributions in a hydraulic excavator control application with hardware experiments.
To this end, we train LMNs from recorded, noisy data and derive feedforward controllers used as part of a leveling assistance system on the excavator.
In our experiments, incorporating disturbance signals and multiple inputs and outputs enhances tracking performance of the learned controller.
A video of our experiments is available at \videourl.
\end{abstract}

\section{INTRODUCTION}

Learning-based and data-driven methods for analyzing and designing control systems using collected system data are of growing interest~\cite{hou2013model,brunke2022}, as they reduce or eliminate the need for time-consuming and sometimes inaccurate first-principles-based system modeling.
This is particularly advantageous in applications where systems are subject to high-mix, low-volume manufacturing, such as mobile working machines for mining and construction, and custom industrial automation solutions.
The classic indirect, data-driven control approach involves performing system identification~\cite{oliver2020nonlinear, lennart1999system} followed by controller synthesis based on the identified system~\cite{kaiser2018sparse, kaiser2021data}. For nonlinear systems, there are various model architectures to choose from, such as neural networks or Gaussian processes. While these models often have high modeling capacity, they can be difficult to interpret or derive controllers from due to the many (hyper)parameters involved.
Additionally, large models may exhibit unfavorable extrapolation behavior and can be challenging to evaluate in real-time on embedded hardware~\cite{rabenstein_2022}.
For some of these shortcomings, \glspl{lmn} offer a promising alternative~\cite{oliver2020nonlinear} .
\Glspl{lmn} can model nonlinear system behavior, show favorable extrapolation behavior~\cite{SCHUSSLER2021487}, and can even be adapted online~\cite{hametner_2013}, while their simple structure allows for interpretability~\cite{murray1994local}.
For \gls{siso} systems, feedforward controllers can be automatically derived from \glspl{lmn} through feedback linearization~\cite{hametner_2014,rolle_2016}. 
However, this method was previously restricted to \glspl{lmn} without \gls{zd}, which substantially limited the model structure and, consequently, its applications.
In this paper, we use feedback linearization to automatically generate feedforward controllers from \glspl{lmn}. The \glspl{lmn} are trained on real world data from a hydraulic excavator, and the resulting policies are used to control the velocity of its cylinders in a trajectory tracking application (see Fig.~\ref{fig:controlCircuit}).

\newcommand{\gts}{\footnotesize}
\newsavebox{\trainVelPlot}
\savebox{\trainVelPlot}{%
    \tikzset{external/optimize=false}%
    \includegraphics{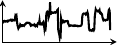}
}
\newsavebox{\trainJoyPlot}
\savebox{\trainJoyPlot}{%
    \tikzset{external/optimize=false}%
    \includegraphics{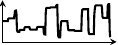}
}
\definecolor{colData}{HTML}{33a02c}
\colorlet{colDataLight}{colData!10}
\definecolor{colLMN}{HTML}{e31a1c}
\colorlet{colLMNLight}{colLMN!10}
\definecolor{colFF}{HTML}{1f78b4}
\colorlet{colFFLight}{colFF!10}
\begin{figure}[t!]
    \centering
    \includegraphics{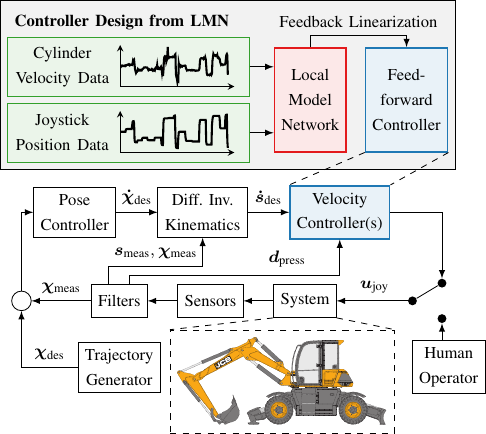}
    \caption{Controller design for and block diagram of the trajectory tracking control on a hydraulic excavator~\cite{jcb_sales_limited_jcb_2018}, the overall system structure is similar to~\cite{rabenstein_2022,weigand_2021}. This paper focuses on the controller design by learning local model networks~{\color{colLMN}(red)} from data {\color{colData} (green)} and using feedback linearization to automatically derive feedforward velocity controllers~{\color{colFF}(blue)}.}
    \label{fig:controlCircuit}
\end{figure}

Recent research~\cite{rabenstein_2022,weigand_2021,egli_2020} has focused on developing assistance functions for hydraulic excavators (e.g., for leveling tasks) that are not yet available in commercial products, with the goal of reducing workload and increasing productivity.
Modeling the dynamics of excavator hydraulics using first principles is challenging.
The highly nonlinear system behavior further complicates precise control—leveling with centimeter accuracy, for instance, requires years of experience for human expert operators to master.
Moreover, hydraulic excavators are designed to be robust and long-lasting, with limited sensors and computing resources~\cite{jiang2020overview}.
This constrains the range of applicable methods and makes controller design a challenging task.
Pure linear feedback controllers show poor tracking in practice. Therefore, this work focuses on the design of nonlinear feedforward controllers for the hydraulics.
The high-mix, low-volume manufacturing of hydraulic excavators makes them ideal candidates for data-driven control approaches.
Yet the interpretability of learned models is paramount to ensure safe operation.
Controllers derived from learned \glspl{lmn} are well suited to the challenges of excavator control, which will be addressed throughout this paper.
We further extend the existing controller design method by addressing some of its limitations.
In summary, we make the following contributions:
\begin{compactenum}
    \item We derive a criterion for when the controller design from~\cite{rolle_2016} can be applied to \glspl{lmn} with \gls{zd} for the case where the relative degree is one~(Sec.~\ref{sec:contrStab}).
    \item\label{contrib:2} We extend the controller design from \glspl{lmn} via feedback linearization to \gls{mimo} systems~(Sec.~\ref{sec:coupledSystems}).
    \item\label{contrib:3} We extend the controller design to compensate measured disturbances~(Sec.~\ref{sec:disturbanceCompensation}).
    \item We apply this approach in a trajectory tracking application on a real hydraulic excavator~(Sec.~\ref{sec:experiments}). In our experiments, controllers with~\ref{contrib:2}) and~\ref{contrib:3}) outperform the baseline from~\cite{rolle_2016}.
\end{compactenum}

\subsection{Related Work}

In this section, we discuss related work on feedforward controller design from \glspl{lmn} via feedback linearization. We will also present some existing approaches to data-driven control for hydraulic excavators, which is the main application in our experiments.

\subsubsection{Feedback Linearization of Local Model Networks}
Automatic derivation of feedforward controllers from \glspl{lmn} is first introduced in~\cite{hametner_2014}.
For controller synthesis via feedback linearization in general, the \gls{zd} (cf.~\cite{isidori2013zero}) of the model need to be stable if they exist. 
Since the feedforward controller has internal feedback, unstable \gls{zd} may lead to an unbounded growth of the controller output and it becomes unusable~\cite{isidori_nonlinear_1995}.
In \cite{rolle_2016} a criterion for the existence of the \gls{zd} is derived in the context of~\glspl{lmn}, however, a criterion for the stability is not yet available.
Therefore, the controller design in~\cite{rolle_2016} can only be applied to \glspl{lmn} without \gls{zd}. 
In our work, we overcome this limitation and avoid potentially unbounded outputs of the feedforward controller.
We propose to use the \gls{bibo} stability of the feedforward controller directly as a criterion for when feedback linearization can be applied to \glspl{lmn} with \gls{zd}.

\subsubsection{Data-driven Control for Hydraulic Excavators}
Several recent works address the trajectory tracking problem for hydraulic excavators using data-driven control~\cite{rabenstein_2022,weigand_2021,egli_2020}.
In~\cite{egli_2020}, collected machine data is used to train a \gls{mlp} representing the hydraulics of the system.
Using this model in a simulation environment, a policy (also a \gls{mlp}) is trained using reinforcement learning.
This yields a well performing controller.
However, the training process requires long computation times and the resulting black box model and controller lack interpretability, making the safety of the system difficult to certify.

Instead of the black-box \gls{mlp} approach, a differential inverse kinematics model can be employed to focus the controller design task on the hydraulics~\cite{rabenstein_2022,weigand_2021}, which we also do in our application.
The lower half of Fig.~\ref{fig:controlCircuit} shows the overall control structure that we use for the trajectory tracking, which is similar to the structure used in~\cite{rabenstein_2022,weigand_2021}. The desired \gls{tcp} position $\excDesTcp$, provided by the trajectory generator, is compared to the current \gls{tcp} position $\excMeasTcp$, estimated from the measured cylinder positions $\excMeasCyl$, to generate a desired \gls{tcp} velocity $\excDesTcpVel$.
Using the differential inverse kinematics model, this is mapped to the desired cylinder velocities $\excDesCylVel$.
The velocity controller generates the control signals $\excJoyPos$ from the desired cylinder velocities and measured pressure $\excPress$.
These signals are the same that would be controlled by an operator using two joysticks and are thus referred to as joystick signals in the following.
Each signal controls the valves corresponding to one of the hydraulic cylinders.
Notably, the control performance of the cylinders is not independent as they are coupled through the kinematics and hydraulics.

Existing methods to learn cylinder velocity controllers rely on \glspl{mlp}~\cite{weigand_2021} or \mbox{\gls{gpr}~\cite{rabenstein_2022}.}
Both works require a first principles model for improving extrapolation behavior. %
Instead, we solely rely on system identification using \glspl{lmn} to synthesize velocity controllers using feedback linearization. 
The good extrapolation behavior of \glspl{lmn} allows us to avoid time-consuming first principles modelling. While in~\cite{rabenstein_2022} the \gls{gpr} has to be adapted in order to be used in real-time, the simple structure of the \gls{lmn} and the resulting controller allow for fast inference.
Compared to~\cite{egli_2020}, our model structure improves interpretability.

\section{PRELIMINARIES}

In this section, we introduce the structure of \glspl{lmn} and how they can be used to describe nonlinear dynamic systems. We will further introduce the transformation of \glspl{lmn} to state-space representation as well as the resulting feedforward control law derived in~\cite{rolle_2016}, which we base our subsequent contributions on.

\subsection{Problem Setting}

We consider a stable, continuous, real world system 
\begin{equation}\label{eq:realSys}
    \begin{aligned}
        \sysStateChange &= \sysSysEq{\sysState,\sysInp,\sysDist}, \\
        \sysOutp &= \sysOutpEq{\sysState,\sysInp,\sysDist}
    \end{aligned}
\end{equation}
with unknown dynamics, inputs $\sysInp$, outputs $\sysOutp$, and disturbance $\sysDist$. 
\begin{assumption}
    The inputs $\sysInp$, outputs $\sysOutp$, and disturbance $\sysDist$ can be measured and $\sysDist$ is independent of $\sysInp$.
\end{assumption}

In our application, this system corresponds to the hydraulics of the excavator cylinders, with the joystick signals as the inputs and the cylinder velocities as the outputs.

\subsection{Local Model Networks}
\Glspl{lmn} are a special type of neural network that approximate nonlinear functions using a combination of \glspl{llm}~\cite{oliver2020nonlinear}. The \gls{lmn} output ${\lmnOutp \in \R}$ is calculated as the weighted sum of the \gls{llm} outputs ${\lmnModelOutp{\lmnModelIdx} \in \R}$ as 
\begin{align}
    \lmnOutp &= \textstyle\sum_{\lmnModelIdx=1}^\lmnNumModels \lmnModelOutp{\lmnModelIdx} \lmnModelAct{\lmnModelIdx}, \label{eq:lmn_output} \\
    \lmnModelOutp{\lmnModelIdx} &= \lmnModelWeight{\lmnModelIdx}{0} + \lmnModelWeight{\lmnModelIdx}{1}\lmnLinInp{1} + \hdots + \lmnModelWeight{\lmnModelIdx}{\lmnLinInpSize}\lmnLinInp{\lmnLinInpSize},
\end{align}
where the weights ${\lmnModelAct{\lmnModelIdx} \in [0,1] \subset \R}$ are the so-called validity and $\lmnModelWeight{\lmnModelIdx}{\genIdx}$ are the parameters of the \glspl{llm}.
The number of local models is given as $\lmnNumModels \in \N^+$. 
An important aspect of \glspl{lmn} is that its two input vectors, ${\lmnLinInpVec \in \R^\lmnLinInpSize}$ to calculate the \gls{llm} outputs and ${\lmnActInpVec \in \R^{\lmnActInpSize}}$ to calculate the validities, can be treated individually. This is especially important when dealing with dynamic processes, where delayed versions of physical inputs are strongly correlated and should only be incorporated in $\lmnLinInpVec$, but are omitted in $\lmnActInpVec$. 
The validity functions determine in which input region a \gls{llm} is valid. They are normalized, such that 
\begin{equation}\label{eq:validity_fncs}
    \lmnModelAct{\lmnModelIdx} = \frac{\lmnModelVal{\lmnModelIdx}}{\sum_{\lmnModelIdx=1}^{\lmnNumModels}\lmnModelVal{\lmnModelIdx}}
    \text{ with }
    \textstyle\sum_{\lmnModelIdx=1}^{\lmnNumModels}\lmnModelAct{\lmnModelIdx}=1,
\end{equation}
where $\lmnModelVal{\lmnModelIdx}$ is usually chosen as an axis-orthogonal Gaussian, leaving the center and standard deviation for each dimension in $\lmnActInpVec$ as parameters.
The described structure is similar to \gls{tsfm}~\cite{takagi1985fuzzy} and under certain assumptions they are equivalent~\cite{oliver2020nonlinear}.

In this work, we utilize the \gls{lolimot} algorithm with local estimation to fit the parameters for the \glspl{llm} and the validity functions~\cite{oliver2020nonlinear}. This is done by gradually partitioning the input space with axis-orthogonal splits.

In order to use a \gls{lmn} as a discrete-time dynamic model to represent~\eqref{eq:realSys}, it can be used in a \gls{narx} setup~\cite{oliver2020nonlinear}.
For a \gls{siso} system with disturbance, the two input vectors for each time step are 
\begin{equation}\label{eq:narx}
    \narxOutp[\timestep+1] = \narxLmn,
\end{equation}
\begin{align}
    \narxLinInpVec &= \begin{bNiceArray}{c@{}c!{\hspace\tabcoldots}c!{\hspace\tabcoldots}c@{}c}
        [ & \narxInp[\timestep-\narxInpLinDel{1}+1] & \hdots & \narxInp[\timestep-\narxInpLinDel{|\narxInpLinDelSet|}+1] & ]^\top \\
        [ & \narxDist[\timestep-\narxDistLinDel{1}+1] & \hdots & \narxDist[\timestep-\narxDistLinDel{|\narxDistLinDelSet|}+1] & ]^\top \\
        [ & \narxOutp[\timestep-\narxOutpLinDel{1}+1] & \hdots & \narxOutp[\timestep-\narxOutpLinDel{|\narxOutpLinDelSet|}+1] & ]^\top
    \end{bNiceArray}, \\
    \narxActInpVec &= \begin{bNiceArray}{c@{}c!{\hspace\tabcoldots}c!{\hspace\tabcoldots}c@{}c}
        [ & \narxInp[\timestep-\narxInpActDel{1}+1] & \hdots & \narxInp[\timestep-\narxInpActDel{|\narxInpActDelSet|}+1] & ]^\top \\
        [ & \narxDist[\timestep-\narxDistActDel{1}+1] & \hdots & \narxDist[\timestep-\narxDistActDel{|\narxDistActDelSet|}+1] & ]^\top \\
        [ & \narxOutp[\timestep-\narxOutpActDel{1}+1] & \hdots & \narxOutp[\timestep-\narxOutpActDel{|\narxOutpActDelSet|}+1] & ]^\top
    \end{bNiceArray}.
\end{align}
The delays for the input, disturbance, and feedback are given as ${\narxInpLinDel{\genIdx} \in \narxInpLinDelSet}$, ${\narxDistLinDel{\genIdx} \in \narxDistLinDelSet}$, and ${\narxOutpLinDel{\genIdx} \in \narxOutpLinDelSet}$ for the \gls{llm} inputs, respectively. 
For the inputs to the validity, the delays are given as ${\narxInpActDel{\genIdx} \in \narxInpActDelSet}$, ${\narxDistActDel{\genIdx} \in \narxDistActDelSet}$, and ${\narxOutpActDel{\genIdx} \in \narxOutpActDelSet}$.
The delay sets ${\narxInpLinDelSet,\narxDistLinDelSet,\narxOutpLinDelSet,}{\narxInpActDelSet,\narxDistActDelSet,\narxOutpActDelSet\subset\N^+}$ are hyperparameters with ${\narxInpActDelSet \subseteq \narxInpLinDelSet}$, ${\narxDistActDelSet \subseteq \narxDistLinDelSet}$, and ${\narxOutpActDelSet \subseteq \narxOutpLinDelSet}$.
The maximum delays are defined as ${\narxMaxInpDel=\max(\narxInpLinDelSet)}$, ${\narxMaxDistDel=\max(\narxDistLinDelSet)}$, and ${\narxMaxOutpDel=\max(\narxOutpLinDelSet)}$.
Similar to~\eqref{eq:narx}, a \gls{mimo} model can be represented in a \gls{narx} setup by using one \gls{lmn} per output~\cite{oliver2020nonlinear}.

Using a series of discrete observations ${(\sysInp[\sysTime_\timestep],\sysOutp[\sysTime_\timestep],\sysDist[\sysTime_\timestep])}$ from~\eqref{eq:realSys} at a constant sampling rate, the parameters of~\eqref{eq:narx} can be trained using the \gls{lolimot} algorithm to obtain a system model~(cf.~\cite{oliver2020nonlinear}).

\subsection{Feedback Linearization of Local Model Networks}\label{sec:feedbackLinearization}
To apply feedback linearization to a \gls{lmn} as shown in~\eqref{eq:narx}, it must first be transformed into a state-space representation~\cite{henson_1997}.
In this section, we present the transformation and the control law derived in~\cite{hametner_2014} for the case without measured disturbances.

The chosen state-space representation of~\eqref{eq:narx} with ${\narxDistLinDelSet=\narxDistActDelSet=\emptyset}$ has the form of the \gls{lpv} system 
\begin{equation}\label{eq:stateSpace}
    \begin{aligned}
        \dlmnStateVec[\timestep+1] &= \dlmnCombSysMat \dlmnStateVec + \dlmnCombInpMat \dlmnInp + \dlmnCombOffVec, \\
        \dlmnOutp &= \dlmnOutpMat \dlmnStateVec
    \end{aligned}
\end{equation}
with $\dlmnNumStates=\narxMaxInpDel+\narxMaxOutpDel-1$ states. 
We define the vector containing the \gls{llm} validities $\dlmnAct{\lmnModelIdx}$ as $\dlmnActVec=\lmnActVec[\narxActInpVec]$.
The state vector $\dlmnStateVec \in \R^\dlmnNumStates$ is chosen as
\begin{equation}\label{eq:state}
    \dlmnStateVec = \begin{bNiceArray}{c@{}ccc@{}c}
        [ & \dlmnInp[\timestep-\narxMaxInpDel+1] & \hdots & \dlmnInp[\timestep-1] & ]^\top \\
        [ & \dlmnOutp[\timestep-\narxMaxOutpDel+1] & \hdots & \dlmnOutp[\timestep] & ]^\top
    \end{bNiceArray}.
\end{equation}
Since the output depends on past input and output values, these are incorporated into the state.
The system matrix~$\dlmnCombSysMat$, the input matrix~$\dlmnCombInpMat$, and the offset term~$\dlmnCombOffVec$ 
\begin{align}
    \dlmnCombSysMat &= \textstyle\sum_{\lmnModelIdx=1}^\lmnNumModels\dlmnAct{\lmnModelIdx}\dlmnModelSysMat{\lmnModelIdx}
    ,\quad
    \dlmnCombSysMat \in \R^{\dlmnNumStates \times \dlmnNumStates}, \label{eq:lmnCombSysMat}\\
    \dlmnCombInpMat &= \textstyle\sum_{\lmnModelIdx=1}^\lmnNumModels\dlmnAct{\lmnModelIdx}\dlmnModelInpMat{\lmnModelIdx}
    ,\quad
    \dlmnCombInpMat \in \R^{\dlmnNumStates \times 1}, \label{eq:lmnCombInpMat}\\
    \dlmnCombOffVec &= \textstyle\sum_{\lmnModelIdx=1}^\lmnNumModels\dlmnAct{\lmnModelIdx}\dlmnModelOffVec{\lmnModelIdx}
    ,\quad
    \dlmnCombOffVec \in \R^{\dlmnNumStates} \label{eq:lmnCombOffVec}
\end{align}
are defined as the weighted sum of the constant \gls{llm} parameter matrices with the corresponding model validity~$\dlmnAct{\lmnModelIdx}$.
The matrices $\dlmnModelSysMat{\lmnModelIdx}$, $\dlmnModelInpMat{\lmnModelIdx}$, and $\dlmnModelOffVec{\lmnModelIdx}$ are defined according to~\cite[Eq.~5,~10,~11]{rolle_2016}.
The last row of the system matrix $\dlmnModelSysMat{\lmnModelIdx}$ contains the linear parameters 
\begin{align}
    \dlmnModelInpParamTVec{\lmnModelIdx} &= \begin{bmatrix}
        \dlmnModelInpParam{\narxMaxInpDel}{\lmnModelIdx} & \hdots & \dlmnModelInpParam{2}{\lmnModelIdx} \\
    \end{bmatrix},  \\
    \dlmnModelOutpParamTVec{\lmnModelIdx} &= \begin{bmatrix}
        \dlmnModelOutpParam{\narxMaxOutpDel}{\lmnModelIdx} & \hdots & \dlmnModelOutpParam{1}{\lmnModelIdx} \\
    \end{bmatrix}
\end{align}
with $\dlmnModelInpParam{\dlmnInpDel}{\lmnModelIdx}$ as the factor of \gls{llm}~$\lmnModelIdx$ for the input delayed by $\dlmnInpDel$ time steps, $\dlmnModelOutpParam{\dlmnOutpDel}{\lmnModelIdx}$ as the factor for the feedback delayed by $\dlmnOutpDel$ time steps, and $\dlmnModelOffParam{\lmnModelIdx}$ as an offset term. 
Further, $\dlmnModelInpParam{\dlmnInpDel}{\lmnModelIdx} = 0$ if $\dlmnInpDel \not \in \narxInpLinDelSet$ and $\dlmnModelOutpParam{\dlmnOutpDel}{\lmnModelIdx} = 0$ if $\dlmnOutpDel \not \in \narxOutpLinDelSet$.
The constant output matrix is given as 
\begin{equation}
    \dlmnOutpMat = 
    \begin{bmatrix}
        \bm{0}_{1 \times (\dlmnNumStates-1)} & 1 \\
    \end{bmatrix}
    ,\quad
    \dlmnOutpMat \in \R^{1 \times \dlmnNumStates}.
\end{equation}
In the following, we will use 
$\dlmnCombInpParam{\dlmnInpDel} = \textstyle\sum_{\lmnModelIdx=1}^\lmnNumModels\dlmnAct{\lmnModelIdx}\dlmnModelInpParam{\dlmnInpDel}{\lmnModelIdx}$, 
$\dlmnCombOutpParam{\dlmnOutpDel} = \textstyle\sum_{\lmnModelIdx=1}^\lmnNumModels\dlmnAct{\lmnModelIdx}\dlmnModelOutpParam{\dlmnOutpDel}{\lmnModelIdx}$, and 
$\dlmnCombOffParam = \textstyle\sum_{\lmnModelIdx=1}^\lmnNumModels\dlmnAct{\lmnModelIdx}\dlmnModelOffParam{\lmnModelIdx}$
as notation
for the linear parameters weighted with their respective validity.

For feedback linearization, the relative degree $\relDeg$ of a system needs to be known.
We refer to the definition of a well defined relative degree in~\cite[Ch.~4]{henson_1997}.
The control law can be derived from the feedback linearization of~\eqref{eq:stateSpace}, given as 
\begin{equation}\label{eq:invLmnContrLaw}
    \begin{aligned}
        \dlmnDes[\timestep+\relDeg]=\dlmnOutpMat[&\overset{\curvearrowleft}{\textstyle\prod}{}_{\genDel=0}^{\relDeg-1}\dlmnCombSysMat[\timestep+\genDel]\dlmnStateVecDes \\
        +&\overset{\curvearrowleft}{\textstyle\prod}{}_{\genDel=1}^{\relDeg-1}\dlmnCombSysMat[\timestep+\genDel]\dlmnCombInpMat\dlmnInp \\
        +&\textstyle\sum_{j=0}^{\relDeg-1}\overset{\curvearrowleft}{\textstyle\prod}{}_{\genDel=j+1}^{\relDeg-1}\dlmnCombSysMat[\timestep+\genDel] \dlmnCombOffVec[\timestep+j]].
    \end{aligned}
\end{equation}
We use notation $\overset{\curvearrowleft}{\prod}{}^n_{j=1}M_j=M_nM_{n-1}\cdots M_1$ to avoid ambiguity.
The vector $\dlmnStateVecDes$ is chosen as $\dlmnStateVec$ with ${\dlmnOutp=\dlmnDes}$ to 
\begin{equation}\label{eq:dlmnStateVecDes}
    \dlmnStateVecDes = \begin{bNiceArray}{c@{}ccc@{}c}
        [ & \dlmnInp[\timestep-\narxMaxInpDel+1] & \hdots & \dlmnInp[\timestep-1] & ]^\top \\
        [ & \dlmnDes[\timestep-\narxMaxOutpDel+1] & \hdots & \dlmnDes[\timestep] & ]^\top
    \end{bNiceArray}.
\end{equation}
By replacing the desired output with ${\dlmnDes[\timestep]=\dlmnDesShift[\timestep-\relDeg-1]}$ as the new time-shifted desired output, the feedforward controller no longer achieves exact tracking, but results in a causal system~\cite{hametner_2014}. 
For $\narxInpActDelSet = \emptyset$ the \gls{lmn} is input-affine and~\eqref{eq:invLmnContrLaw} can be solved explicitly for $\dlmnInp$.
Otherwise, the validity vector $\dlmnActVec$ may depend on $\dlmnInp$, requiring a numerical solution.

\section{PERMISSIBILITY OF FEEDBACK LINEARIZATION}\label{sec:contrStab}

Feedback linearization requires the \gls{zd} of a system -- in our case, a learned LMN -- to be stable, if they exist~\cite{henson_1997}.
Otherwise, the output of the resulting feedforward controller might grow unbounded. 
In \cite{rolle_2016}, a criterion based on the delays used was provided to ensure the absence of \gls{zd} and that feedback linearization can be applied.
In the first part of this section, we present a counterexample demonstrating that the criterion in~\cite{rolle_2016} is not always sufficient and may result in unbounded controller output. 
Subsequently, we provide a novel criterion for the \gls{bibo} stability of the resulting feedforward controller~\eqref{eq:invLmnContrLaw} specifically for the case where the relative degree is one.
This new criterion, formulated as a linear matrix inequality, considers the \gls{lmn} parameters, specifically the linear ones.
It can be used directly to determine if the controller design presented in {Section~\ref{sec:feedbackLinearization}} can be applied to a \gls{lmn} as~\eqref{eq:stateSpace} or if unstable \gls{zd} may lead to an unbounded controller output due to the controller's internal feedback.

\subsection{Criterion for Existence of Zero Dynamics}
The \gls{zd} of a system only exist, if its relative degree is smaller than the system order~\cite{henson_1997}.
Previous work derived the relative degree for a \gls{lmn} to be 
\begin{equation}\label{eq:lmnRelDegGeneral}
    \relDeg = \min(\narxInpLinDelSet \cup \narxInpActDelSet)
\end{equation}
in general~\cite[Eq.~39]{rolle_2016}.
However, there are multiple cases, where the relative degree is larger or not well defined locally.
For example, \eqref{eq:lmnRelDegGeneral} additionally requires ${\dlmnCombInpParam{\relDeg} \not = 0}$ if ${\relDeg \notin \narxInpActDelSet}$ and therefore ${\dlmnModelInpParam{\relDeg}{\lmnModelIdx}}$ to have the same sign and be nonzero for all $\lmnModelIdx \in [\lmnNumModels]$ with the notation $[\lmnNumModels]=\{1,\hdots,\lmnNumModels\}$.
Otherwise, $\dlmnCombInpParam{\relDeg}$ can become zero locally and would result in a not well defined relative degree.

According to~\cite{rolle_2016}, there are no \gls{zd} for $|\narxInpLinDelSet|=1$ and $\narxInpActDelSet\subseteq\narxInpLinDelSet$ and therefore the feedback linearization can be performed.
The following example shows that this not always true.

\begin{counterexample}
    Choosing $\narxInpLinDelSet = \{2\}$, $\narxOutpLinDelSet = \{1,2\}$, and $\narxInpActDelSet = \emptyset$ fulfills the given condition. We assume parameters are chosen, such that the relative degree $\relDeg=2$ is well defined. With a single local model $\lmnNumModels=1$, the validity $\dlmnAct{1}=1$ is constant. Choosing $\dlmnModelOffParamNoMod=0$ results in a linear time-invariant system. The control law \eqref{eq:invLmnContrLaw} can be written as the transfer function 
    \begin{equation}
        \frac{\dlmnInpZ}{\dlmnDesShiftZ} = \frac{\discrVar^3-(\dlmnModelOutpParamNoMod{1}^2+\dlmnModelOutpParamNoMod{2})\discrVar-\dlmnModelOutpParamNoMod{1}\dlmnModelOutpParamNoMod{2}}{\dlmnModelInpParamNoMod{2}\discrVar^2+\dlmnModelOutpParamNoMod{1}\dlmnModelInpParamNoMod{2}\discrVar}
    \end{equation}
    with a critical pole  of $\pole{2}=-\dlmnModelOutpParamNoMod{1}$, which is only stable for $-1 < \dlmnModelOutpParamNoMod{1} < 1$. Therefore, the controller might have an unbounded output signal for some inputs.
\end{counterexample}

This counterexample shows that a criterion for the existence of \gls{zd} based solely on the delays used must be even more restrictive than the one proposed in~\cite{rolle_2016}. 
For the feedback linearization of a \gls{lmn}, it is therefore reasonable to also consider its parameters to avoid overly restricting its structure.

\subsection{A Novel, Parameter Based Criterion}
In this section, we derive a novel criterion for the permissibility of feedback linearization to \glspl{lmn}.

\begin{definition}
    Following~\cite{chen2004stability}, we define the feedforward controller~\eqref{eq:invLmnContrLaw} as \gls{bibo} stable if
    $ \forall \dlmnDesShift,\dlmnInp \in \R $ there exist $ \genConst_v,\genConst_u \in \R$ such that $||\dlmnDesShift|| < \genConst_v\Rightarrow||\dlmnInp|| \le \genConst_u$.
\end{definition}

In the following, we will focus on the case where the relative degree is one. 

\begin{assumption}\label{ass:relDeg1}
    The dynamic \gls{lmn}~\eqref{eq:stateSpace} has a well defined relative degree $\relDeg=1$ in $(x_0,u_0) \;\forall x_0 \in \R^\dlmnNumStates, u_0 \in \R$.
\end{assumption}

Using Assumption~\ref{ass:relDeg1}, the control law can be written as 
\begin{equation}\label{eq:invLmnContrLawRelDeg1}
    \begin{aligned}
        \dlmnDesShift[\timestep-1] &= \dlmnCombInpParam{1}\dlmnInp \\
        &\phantom{={}} + 
             [\dlmnCombInpParamTVec, \dlmnCombOutpParamTVec]
        \dlmnStateVecDesShift 
        +\dlmnCombOffParam,
    \end{aligned}
\end{equation}
where the vectors 
$\dlmnCombInpParamTVec = [\dlmnCombInpParam{\narxMaxInpDel}, \hdots, \dlmnCombInpParam{2}]$ and 
$\dlmnCombOutpParamTVec = [\dlmnCombOutpParam{\narxMaxOutpDel}, \hdots, \dlmnCombOutpParam{1}]$ 
contain linear parameters weighted with their validities. 
The vector $\dlmnStateVecDesShift$ is $\dlmnStateVec$ with $\dlmnOutp=\dlmnDesShift[\timestep-2]$, similarly to \eqref{eq:dlmnStateVecDes}. 
We split \eqref{eq:invLmnContrLawRelDeg1} into a stable input scheduler and a \gls{lpv} system following~\cite{mayr_2021}. 
The \gls{lpv} state-space representation 
\begin{align}
    \invLmnStateVec[\timestep+1] &= \invLmnCombSysMat\invLmnStateVec + \invLmnCombInpMat \invLmnDesShiftVec + \invLmnCombOffVec, \label{eq:invLmnStateSpace} \\
    \dlmnInp &= \invLmnOutpMat\invLmnStateVec
\end{align}
has $\invLmnNumStates=\narxMaxInpDel-1$ states with the vectors
\begin{align}
    \invLmnStateVec &= \begin{bmatrix}
        \dlmnInp[\timestep-\narxMaxInpDel+2] & \hdots & \dlmnInp
    \end{bmatrix}^\top, \\
    \invLmnDesShiftVec &= \begin{bmatrix}
        \dlmnDesShift[\timestep-\narxMaxOutpDel] & \hdots & \dlmnDesShift
    \end{bmatrix}^\top.
\end{align}
The system matrix ${\invLmnCombSysMat \in \R^{\invLmnNumStates \times \invLmnNumStates}}$, input matrix ${\invLmnCombInpMat \in \R^{\invLmnNumStates \times \narxMaxOutpDel+1}}$ and the offset term ${\invLmnCombOffVec \in \R^{\invLmnNumStates \times 1}}$ are 
\begin{align}
    \invLmnCombSysMat &= \invLmnCombFacMat \textstyle\sum_{\lmnModelIdx=1}^{\lmnNumModels} \dlmnAct{\lmnModelIdx} \invLmnModelSysMat{\lmnModelIdx} \label{eq:invLmnCombSysMat}, \\
    \invLmnCombInpMat &= \invLmnCombFacMat \textstyle\sum_{\lmnModelIdx=1}^{\lmnNumModels} \dlmnAct{\lmnModelIdx} \invLmnModelInpMat{\lmnModelIdx}, \\
    \invLmnCombOffVec &= \invLmnCombFacMat \textstyle\sum_{\lmnModelIdx=1}^{\lmnNumModels} \dlmnAct{\lmnModelIdx} \invLmnModelOffVec{\lmnModelIdx},
\end{align}
which again can be represented as weighted sums of the constant matrices
\begin{equation}
    \invLmnModelSysMat{\lmnModelIdx} = \begin{bmatrix}
        \begin{bmatrix}
            \bm{0}_{\narxMaxInpDel-2 \times 1} & \bm{I}_{\narxMaxInpDel-2}
        \end{bmatrix} \\
        \dlmnModelInpParamTVec{\lmnModelIdx}
    \end{bmatrix},
\end{equation}
\begin{tabularx}{\columnwidth}{@{}XX@{}}
    \vspace{-\baselineskip}\begin{equation}
        \invLmnModelInpMat{\lmnModelIdx} = \begin{bmatrix}
            \bm{0}_{\narxMaxInpDel-2 \times \narxMaxOutpDel+1} \\
            \begin{bmatrix}
                \dlmnModelOutpParamTVec{\lmnModelIdx} & 1
            \end{bmatrix}
        \end{bmatrix},
    \end{equation} & 
    \vspace{-\baselineskip}\begin{equation}
        \invLmnModelOffVec{\lmnModelIdx} = \begin{bmatrix}
            \bm{0}_{\narxMaxInpDel-2 \times 1} \\
            \dlmnModelOffParam{\lmnModelIdx}
        \end{bmatrix}.
    \end{equation}
\end{tabularx}
\vspace{-\baselineskip}\\\noindent
Additionally, there is a validity dependent factor 
\begin{align}
    \invLmnCombFacMat &= \begin{bmatrix}
        \bm{I}_{\narxMaxInpDel-2 \times \narxMaxInpDel-2} & \bm{0}_{\narxMaxInpDel-2 \times 1} \\
        \bm{0}_{1 \times \narxMaxInpDel-2} & \invLmnCombFac
    \end{bmatrix}, \\
    \invLmnCombFac &= -\frac{1}{\dlmnCombInpParam{1}}
\end{align}
for the last rows of $\invLmnCombSysMat$, $\invLmnCombInpMat$, and $\invLmnCombOffVec$.

\begin{assumption}\label{ass:sameSign}
    All $\dlmnModelInpParam{1}{\lmnModelIdx}$ have the same sign and are non-zero for $\lmnModelIdx \in [\lmnNumModels]$.
\end{assumption}

\begin{lemma}\label{lma:stability}
    Let Assumptions~\ref{ass:relDeg1} and~\ref{ass:sameSign} hold. Consider the open-loop dynamics of the feedforward controller \eqref{eq:invLmnStateSpace} 
    \begin{equation}\label{eq:invLmnOpenLoop}
        \invLmnStateVec[\timestep+1] = \invLmnCombSysMat\invLmnStateVec.
    \end{equation}
    Its equilibrium $\invLmnStateVecEq=\bm{0}$ is \gls{gas} if there exists a positive definite matrix $\posDefMat$ such that 
    \begin{align}
        (\invLmnModelFacMat{\lmnModelIdx}\invLmnModelSysMat{\lmnModelIdx})^\top\posDefMat(\invLmnModelFacMat{\lmnModelIdx}\invLmnModelSysMat{\lmnModelIdx})-\posDefMat \prec \bm{0}
        \quad 
        \forall \lmnModelIdx \in [\lmnNumModels], \label{eq:stabilityCriterion} \\
        \invLmnModelFacMat{\lmnModelIdx} = \begin{bmatrix}
            \bm{I}_{\narxMaxInpDel-2} & \bm{0}_{\narxMaxInpDel-2 \times 1} \\
            \bm{0}_{1 \times \narxMaxInpDel-2} & \invLmnModelFac{\lmnModelIdx} \\
        \end{bmatrix},\quad
        \invLmnModelFac{\lmnModelIdx} = -\frac{1}{\dlmnModelInpParam{1}{\lmnModelIdx}}.
    \end{align}
The proof of the lemma is given in Appendix~\ref{apx:proofStability}.
\end{lemma}

\begin{theorem}\label{thm:bibo}
    Let Assumptions~\ref{ass:relDeg1} and~\ref{ass:sameSign} hold. The feedforward controller~\eqref{eq:invLmnContrLawRelDeg1} is \gls{bibo} stable if there exists a positive definite matrix $\posDefMat$ such that \eqref{eq:stabilityCriterion} holds.
\end{theorem}
The proof of the theorem is given in Appendix~\ref{apx:proofBiboStability}

Theorem~\ref{thm:bibo} allows us to apply feedback linearization to \glspl{lmn} with \gls{zd} while guaranteeing some degree of controller stability.
Note that the type of validity function is not specified.
Depending on the type of function, stronger notions of stability may apply.

\section{DISTURBANCE COMPENSATION}\label{sec:disturbanceCompensation}
Real-world systems, such as hydraulic excavators, are often subject to disturbances.
In the following, we derive the feedforward control law for a \gls{lmn}~\eqref{eq:narx} that also compensates disturbances.
We consider disturbance signals as inputs to our model that can be measured online.
In the hydraulic excavator, an example for such a disturbance could be hydraulic pressure fluctuations, which can be measured with sensors.

The state-space representation of a LMN~\eqref{eq:narx} including disturbance is 
\begin{equation}\label{eq:stateSpaceDist}
    \begin{aligned}
        \dlmnStateVecDist[\timestep+1] &= \dlmnCombSysMatDist \dlmnStateVecDist + \dlmnCombInpMatDist \dlmnInp \\
        &\phantom{={}}+ \dlmnCombDistMatDist \dlmnDist + \dlmnCombOffVecDist, \\
        \dlmnOutp &= \dlmnOutpMat \dlmnStateVecDist,
    \end{aligned}
\end{equation}
where the disturbance $\dlmnDist$ enters as a second input. 
There are $\dlmnNumStatesDist=\narxMaxInpDel+\narxMaxDistDel+\narxMaxOutpDel-2$ states and the extended state-space vector $\dlmnStateVecDist \in \R^\dlmnNumStatesDist$ has the form 
\begin{equation}
    \dlmnStateVecDist = \begin{bNiceArray}{c@{}ccc@{}c}
        [ & \dlmnInp[\timestep-\narxMaxInpDel+1] & \hdots & \dlmnInp[\timestep-1] & ]^\top \\
        [ & \dlmnDist[\timestep-\narxMaxDistDel+1] & \hdots & \dlmnDist[\timestep-1] & ]^\top \\
        [ & \dlmnOutp[\timestep-\narxMaxOutpDel+1] & \hdots & \dlmnOutp[\timestep] & ]^\top
    \end{bNiceArray}.
\end{equation}
The system matrix~$\dlmnCombSysMatDist$, the input matrices~$\dlmnCombInpMatDist$ and~$\dlmnCombDistMatDist$, and the offset term~$\dlmnCombOffVecDist$ are defined like \eqref{eq:lmnCombSysMat}-\eqref{eq:lmnCombOffVec} as the linear combination of the constant matrices 
\begin{equation}
    \dlmnModelSysMatDist{\lmnModelIdx} = 
    \left[
    \begin{array}{@{}c|c|c@{}}
        \upperShiftMatrix{\narxMaxInpDel-1} & \bm{0}_{\narxMaxInpDel-1 \times \narxMaxDistDel-1} & \bm{0}_{\narxMaxInpDel-1 \times \narxMaxOutpDel} \\
        \hline & & \\[-1em]
        \bm{0}_{\narxMaxDistDel-1 \times \narxMaxInpDel-1} & \upperShiftMatrix{\narxMaxDistDel-1} & \bm{0}_{\narxMaxDistDel-1 \times \narxMaxOutpDel} \\
        \hline & & \\[-1em]
        \bm{0}_{\narxMaxOutpDel-1 \times \narxMaxInpDel-1} & \bm{0}_{\narxMaxOutpDel-1 \times \narxMaxDistDel-1} & \identitySub{\narxMaxOutpDel} \\
        \dlmnModelInpParamTVec{\lmnModelIdx} & \dlmnModelDistParamTVec{\lmnModelIdx} & \dlmnModelOutpParamTVec{\lmnModelIdx} \\
    \end{array}
    \right],
\end{equation}
\begin{tabularx}{\columnwidth}{@{}XX@{}}
    \vspace{-\baselineskip}\begin{equation}
        \dlmnModelInpMatDist{\lmnModelIdx} = 
        \left[
        \begin{array}{@{}c@{}}
            \bm{0}_{\narxMaxInpDel-2 \times 2} \\
            1 \\
            \hline
            \bm{0}_{\narxMaxDistDel-1 \times 2} \\
            \hline
            \bm{0}_{\narxMaxOutpDel-1 \times 2} \\
            \dlmnModelInpParam{1}{\lmnModelIdx}
        \end{array}
        \right],
    \end{equation} & 
    \vspace{-\baselineskip}\begin{equation}
        \dlmnModelDistMatDist{\lmnModelIdx} = 
        \left[
        \begin{array}{@{}c@{}}
            \bm{0}_{\narxMaxInpDel-1 \times 2} \\
            \hline
            \bm{0}_{\narxMaxDistDel-2 \times 2} \\
            1 \\
            \hline
            \bm{0}_{\narxMaxOutpDel-1 \times 2} \\
            \dlmnModelDistParam{1}{\lmnModelIdx}
        \end{array}
        \right],
    \end{equation} \\
\end{tabularx}
\vspace{-\baselineskip}\\\noindent
\begin{equation}
    \dlmnModelOffVecDist{\lmnModelIdx} = 
    \begin{bmatrix}
        \bm{0}_{\dlmnNumStatesDist-1 \times 1} \\
        \dlmnModelOffParam{\lmnModelIdx}
    \end{bmatrix}
\end{equation}
with the abbreviations $\identitySub{\genDim} \in \R^{\genDim-1 \times \genDim}$ and $\upperShiftMatrix{\genDim} \in \R^{\genDim \times \genDim}$ as 
\begin{tabularx}{\columnwidth}{@{}>{\hsize=1.2\hsize}X>{\hsize=.8\hsize}X@{}}
    \vspace{-\baselineskip}\begin{equation}
        \identitySub{\genDim} = 
        \begin{bmatrix}
            \bm{0}_{\genDim-1 \times 1} & \bm{I}_{\genDim-1}
        \end{bmatrix},
    \end{equation} & 
    \vspace{-\baselineskip}\begin{equation}
        \upperShiftMatrix{\genDim} = 
        \begin{bmatrix}
            \identitySub{\genDim} \\
            \bm{0}_{1 \times \genDim}
        \end{bmatrix}.
    \end{equation} \\
\end{tabularx}
\vspace{-\baselineskip}\\\noindent
The linear parameters for the disturbance are given as 
$\dlmnModelDistParamTVec{\lmnModelIdx} = [
    \dlmnModelDistParam{\narxMaxDistDel}{\lmnModelIdx} \;\hdots \; \dlmnModelDistParam{2}{\lmnModelIdx}
]$
with $\dlmnModelDistParam{\dlmnDistDel}{\lmnModelIdx}$ as the factor of \gls{llm} $\lmnModelIdx$ for the disturbance delayed by $\dlmnDistDel$ time steps. 
The new state-space representation~\eqref{eq:stateSpaceDist} leads to the control law 
\begin{equation}\label{eq:controlLawDist}
    \begin{aligned}
        \dlmnDes[\timestep+\relDeg]=&\dlmnOutpMat[\overset{\curvearrowleft}{\textstyle\prod}{}_{\genDel=0}^{\relDeg-1}\dlmnCombSysMatDist[\timestep+\genDel]\dlmnStateVecDesDist \\
        +&\overset{\curvearrowleft}{\textstyle\prod}{}_{\genDel=1}^{\relDeg-1}\dlmnCombSysMatDist[\timestep+\genDel]\dlmnCombInpMatDist\dlmnInp \\
        +&\textstyle\sum_{\genIdx=0}^{\relDeg-\relDegDist}\overset{\curvearrowleft}{\textstyle\prod}{}_{\genDel=\genIdx+1}^{\relDeg-1} \dlmnCombSysMatDist[\timestep+\genDel] \dlmnCombDistMatDist[\timestep+\genIdx]\dlmnDist[\timestep+\genIdx] \\
        +&\textstyle\sum_{\genIdx=0}^{\relDeg-1}\overset{\curvearrowleft}{\textstyle\prod}{}_{\genDel=\genIdx+1}^{\relDeg-1}\dlmnCombSysMatDist[\timestep+\genDel] \dlmnCombOffVecDist[\timestep+\genIdx]]
    \end{aligned}
\end{equation}
with $\relDegDist$ as the relative degree of the disturbance. 
Note that for $\relDegDist > \relDeg$ only the disturbances already included in the state vector $\dlmnStateVecDesDist$ are used. 
For $\relDegDist \le \relDeg$ future values for $\dlmnDist$ are required. 
From this follows another requirement in order to apply the control law~\eqref{eq:invLmnContrLawRelDeg1}.
\begin{assumption}
    The disturbances $\dlmnDist[\timestep+\genIdx]$ for all ${\genIdx\in\{0,\hdots,\relDeg-\relDegDist\}}$ are known at time step $\timestep-1$ or can be predicted.
\end{assumption}

For a slowly changing $\dlmnDist$ relative to the sampling rate, we can estimate 
$\dlmnDist[\timestep+\genIdx] \approx \dlmnDist[\timestep-1] 
\;
\forall \genIdx \in \{0,\hdots,\relDeg-\relDegDist\}$
as a constant value.
\begin{remark}
    If the disturbance is not independent of the controller output, stable \gls{zd} are not sufficient for the stability of the feedforward control anymore, as feedback is introduced.
\end{remark}

\section{SYSTEMS WITH MULTIPLE INPUTS/OUTPUTS}\label{sec:coupledSystems}
In the previous sections, we considered systems with a single input and a single output. Many real-world systems, such as hydraulic excavators, have multiple inputs and outputs, for example three hydraulic cylinders and associated joystick inputs, which are coupled internally via the kinematics and hydraulics.
In this section, we extend the approach of feedback linearization to derive feedforward controllers to \glspl{lmn} with multiple inputs and outputs.

The state-space representation for a \gls{lmn} in a \gls{narx} setup with multiple inputs and outputs can be written as 
\begin{equation}
    \begin{aligned}
        \dlmnStateVecMimo[\timestep+1] = &\dlmnCombSysMatMimo \dlmnStateVecMimo + \dlmnCombInpMatMimo \dlmnInpVecMimo \\
        & + \dlmnCombOffVecMimo \\
        \dlmnOutpVecMimo = &\dlmnOutpMatMimo \dlmnStateVecMimo.
    \end{aligned}
\end{equation}
For a system of $\dlmnMaxInp$ inputs and $\dlmnMaxOutp$ outputs, this results in a state vector of dimension $\dlmnNumStatesMimo = \textstyle\sum_{\dlmnInpIdx=1}^{\dlmnMaxInp}(\dlmnMaxInpDelMimo{\dlmnInpIdx}-1) + \textstyle\sum_{\dlmnOutpIdx=1}^{\dlmnMaxOutp}\dlmnMaxOutpDelMimo{\dlmnOutpIdx}$.
The maximum delay of input $\dlmnInpMimo{\dlmnInpIdx}$ used across all models is denoted as $\dlmnMaxInpDelMimo{\dlmnInpIdx}$ and the maximum delay of feedback $\dlmnOutpMimo{\dlmnOutpIdx}$ as $\dlmnMaxOutpDelMimo{\dlmnOutpIdx}$. 
Similarly to before, the state vector contains delayed values for the inputs $\dlmnInpMimo{\dlmnInpIdx}$ and the outputs $\dlmnOutpMimo{\dlmnOutpIdx}$ and can be written as
\begin{equation}
    \setlength{\arraycolsep}{\arraycolsepdots}
    \dlmnStateVecMimo = \begin{bNiceArray}{c@{}ccc@{}c}
        [ & \dlmnInpMimo[\timestep-\dlmnMaxInpDelMimo{1}+1]{1} & \hdots & \dlmnInpMimo[\timestep-1]{1} & ]^\top \\
        \multicolumn{5}{c}{\hdots} \\
        [ & \dlmnInpMimo[\timestep-\dlmnMaxInpDelMimo{\dlmnMaxInp}+1]{\dlmnMaxInp} & \hdots & \dlmnInpMimo[\timestep-1]{\dlmnMaxInp} & ]^\top \\
        [ & \dlmnOutpMimo[\timestep-\dlmnMaxOutpDelMimo{1}+1]{1} & \hdots & \dlmnOutpMimo{1} & ]^\top \\
        \multicolumn{5}{c}{\hdots} \\
        [ & \dlmnOutpMimo[\timestep-\dlmnMaxOutpDelMimo{\dlmnMaxOutp}+1]{\dlmnMaxOutp} & \hdots & \dlmnOutpMimo{\dlmnMaxOutp} & ]^\top
    \end{bNiceArray}.
\end{equation}
The validity vector $\dlmnActVecMimoComb$ is a concatenation of the validity vectors of the single \glspl{lmn}. 
The design of matrices $\dlmnCombSysMatMimo$, $\dlmnCombInpMatMimo$, $\dlmnCombOffVecMimo$, and $\dlmnOutpMatMimo$ is similar to \eqref{eq:stateSpaceDist}. 
Note that every input-output pair has a relative degree.
\begin{assumption}\label{ass:relDegSame}
    The relative degree $\relDeg$ is well defined and the same for every input-output pair.
\end{assumption}
Using Assumption~\ref{ass:relDegSame}, the feedforward control law for the \gls{mimo} system is 
\begin{align}
    \dlmnDesVecMimo[\timestep+\relDeg]=&\dlmnOutpMatMimo[\overset{\curvearrowleft}{\textstyle\prod}{}_{\genDel=0}^{\relDeg-1}\dlmnCombSysMatMimo[\timestep+\genDel]\dlmnStateVecDes \nonumber\\
    &+\overset{\curvearrowleft}{\textstyle\prod}{}_{\genDel=1}^{\relDeg-1}\dlmnCombSysMatMimo[\timestep+\genDel]\dlmnCombInpMatMimo\dlmnInpVecMimo \label{eq:invLmnContrLawMimo} \\
    &+\textstyle\sum_{\genIdx=0}^{\relDeg-1}\overset{\curvearrowleft}{\textstyle\prod}{}_{\genDel=\genIdx+1}^{\relDeg-1} \dlmnCombSysMatMimo[\timestep+\genDel] \dlmnCombOffVecMimo[\timestep+\genIdx]].\nonumber
\end{align}
If $\narxInpActDelSet = \emptyset$ for all inputs, the control law yields a system of $\dlmnMaxOutp$ linear equations with $\dlmnMaxInp$ unknowns. 
This can be solved for $\dlmnMaxOutp \le \dlmnMaxInp$ or the quadratic error minimized for $\dlmnMaxOutp > \dlmnMaxInp$.

\section{EXPERIMENTAL RESULTS}\label{sec:experiments}

We demonstrate the effectiveness of training \glspl{lmn} and deriving feedforward controllers using feedback linearization in hardware experiments with a hydraulic excavator.
Based on recorded trajectory data, we train \glspl{lmn} in \gls{narx} structure to predict the hydraulic cylinder velocities. We then derive cylinder velocity controllers based on the methods presented in Sections~\ref{sec:feedbackLinearization}, \ref{sec:disturbanceCompensation}, and \ref{sec:coupledSystems}. We compare the prediction errors of the \glspl{lmn} and the tracking performance of the controllers in hardware experiments.

\begin{figure}[htpb]
    \centering
    \includegraphics{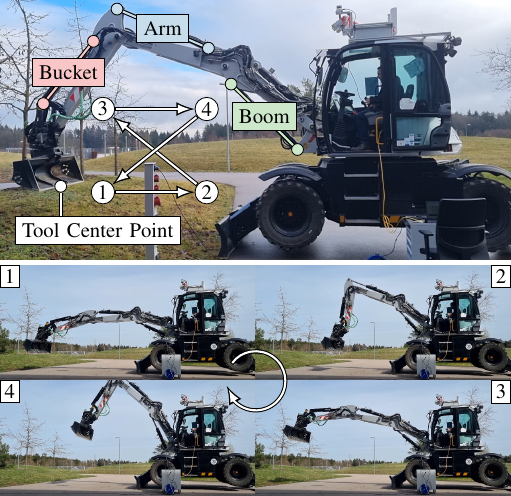}
    \caption{The hydraulic excavator (JCB Hydradig 110W) with controlled hydraulic cylinders and the reference tool center point trajectory tracked in our experiments (top). The configurations of the excavator at the corner points (1-4) of the trajectory are shown at the bottom.}
    \label{fig:experiment}
\end{figure}

The hydraulic excavator used is a JCB Hydradig 110W with three hydraulic cylinders: arm, boom, and bucket (see Fig.~\ref{fig:experiment}).
The cylinder velocity tracking controller therefore has three desired velocities as input and three joystick signals as output.
We evaluate controllers from three types of \glspl{lmn}: 
\begin{enumerate*}[label=(\roman*)]
    \item parallel \gls{siso} models with $\narxInpLinDelSet=\{1\}$ following~\cite{hametner_2014,rolle_2016} and serving as a baseline; \label{itm:siso}
    \item parallel \gls{siso} models with \gls{zd} and $\narxInpLinDelSet=[4]$ fulfilling~\eqref{eq:stabilityCriterion}; and \label{itm:sisoNum}
    \item parallel \gls{mimo} models with $\narxInpLinDelSet=\{1\}$ following Section~\ref{sec:coupledSystems}. \label{itm:mimo}
\end{enumerate*}
For all models we use $\narxOutpLinDelSet=[8]$, $\narxOutpActDelSet=\{1\}$, and $\narxInpActDelSet=\emptyset$, such that the controllers \eqref{eq:invLmnContrLaw}, \eqref{eq:controlLawDist}, and \eqref{eq:invLmnContrLawMimo} admit an explicit form.
We train each model type with and without pressure information, treated as a disturbance following Section~\ref{sec:disturbanceCompensation}.
The pressures used are the load sensing and the pump pressure.
All models are trained on \SI{35}{\minute} and evaluated on \SI{8}{\minute} of trajectory data collected from the excavator using pseudo random joystick signals.
While the data is sampled with \SI{100}{\hertz}, the models use a sampling rate of \SI{6.25}{\hertz}.
The pressure sensor readings are filtered using a first order lowpass, while the cylinder velocity is estimated from the position using a Kalman filter. 
All models are trained using the \gls{lolimot} algorithm with local estimation up to a maximum of 32 \glspl{llm} per \gls{lmn}. 
To ensure model~\ref{itm:sisoNum} fulfills Assumption~\ref{ass:sameSign} and \eqref{eq:stabilityCriterion}, its parameters are fine-tuned using gradient descent.
The accuracies of the trained \glspl{lmn} are tested by a forward prediction and comparison with true, measured data. 
Joystick signals and pressure information from the evaluation data are used as input and previous predictions as feedback.
Table~\ref{tab:simulationRmse} shows the error of the estimated velocity by the \gls{lmn}.
The best-performing \gls{lmn} and controller are highlighted for each cylinder.
Adding pressure information improves the accuracy for all cylinders and every model type. 
This is most prominent for the bucket cylinder with model~\ref{itm:siso}. 
Models~\ref{itm:sisoNum} and~\ref{itm:mimo} show better accuracy than model~\ref{itm:siso} except for the bucket when pressure is used.

\begin{table}[tb]
    \renewcommand{\arraystretch}{1.1}
    \caption{Velocity Prediction Error of LMNs}
    \label{tab:simulationRmse}
    \centering
    \begin{tabular}{c|c|ccc}
         & & \multicolumn{3}{c}{\bf{RMSE [cm/s]}} \\
        \bf{Model Type} & \bf{Pressure} & \bf{Arm} & \bf{Boom} & \bf{Bucket} \\
        \hline
        \multirow{2}{*}{\ref{itm:siso} SISO $\narxInpLinDelSet=\{1\}$} & no & 1.619 & 0.940 & 1.357 \\
         & yes & 1.518 & 0.873 & \bf{0.887} \\
        \hline
        \multirow{2}{*}{\ref{itm:sisoNum} SISO $\narxInpLinDelSet=[4]$} & no & 1.606 & 0.893 & 1.297 \\
         & yes & 1.487 & \bf{0.800} & 1.163 \\
        \hline
        \multirow{2}{*}{\ref{itm:mimo} MIMO $\narxInpLinDelSet=\{1\}$} & no & 1.557 & 0.896 & 1.120 \\
         & yes & \bf{1.487} & 0.871 & 1.074 \\
    \end{tabular}
\end{table}

\begin{table}[phtb]
    \renewcommand{\arraystretch}{1.1}
    \caption{Velocity Tracking Error using Feedforward Controllers}
    \label{tab:excavatorRmse}
    \centering
    \begin{tabular}{c|c|ccc}
        \bf{Controller based on} & & \multicolumn{3}{c}{\bf{RMSE [cm/s]}} \\
        \bf{Model Type} & \bf{Pressure} & \bf{Arm} & \bf{Boom} & \bf{Bucket} \\
        \hline
        \multirow{2}{*}{\ref{itm:siso} SISO $\narxInpLinDelSet=\{1\}$} & no & 1.156 & 0.766 & 1.368 \\
         & yes & 0.945 & 0.523 & \bf{0.776} \\
        \hline
        \multirow{2}{*}{\ref{itm:sisoNum} SISO $\narxInpLinDelSet=[4]$} & no & 0.973 & 0.639 & 0.895 \\
         & yes & \bf{0.861} & 0.611 & 0.939 \\
        \hline
        \multirow{2}{*}{\ref{itm:mimo} MIMO $\narxInpLinDelSet=\{1\}$} & no & 1.034 & 0.582 & 1.078 \\
         & yes & 0.901 & \bf{0.517} & 0.811 \\
    \end{tabular}
\end{table}

The trajectory tracking controllers run on the real excavator using a dSpace MicroAutoBox II.
For evaluation, the desired trajectory was set to an eight-like figure between four points set in the operating space (see Fig.~\ref{fig:experiment} or Video\footnote{\videourl}).
This cycle was repeated twice per controller. Table~\ref{tab:excavatorRmse} shows the velocity tracking error of the cylinders.
Similar to simulation, we see smaller tracking errors when using pressure information in almost every case.
Without pressure, controllers based on models~\ref{itm:sisoNum} and~\ref{itm:mimo} show significantly better performance than~\ref{itm:siso}.
When using pressure, controllers based on models~\ref{itm:sisoNum} and~\ref{itm:mimo} have a noticeable smaller error only in case of the arm.
For the bucket, the controller based on model~\ref{itm:siso} also has the smallest tracking error.
The best performing controller in the case of arm and boom switch when compared to velocity prediction. 
In summary, the extensions of the \gls{siso} model inversion to \gls{mimo}, models with \gls{zd}, and disturbance compensation each provide an improvement, which highlights the importance of all three contributions.
However, in this application, the combination of pressure with \gls{mimo} models or models with \gls{zd} does not further improve performance.
Reasons for this can be, e.g., redundancy of the pressure information when using all joystick signals as inputs.
To determine the repeatability of the experiment, the cycle was repeated six times for the controller from model~\ref{itm:siso} with pressure, resulting in a standard deviation of the RMSE of \SI{0.604}{\milli\metre\per\second}, \SI{0.312}{\milli\metre\per\second}, and \SI{0.297}{\milli\metre\per\second} for arm, boom, and bucket, respectively. 
Fig.~\ref{fig:armVelPlot} shows the desired and the from the measurements estimated cylinder velocities using the controller from model~\ref{itm:siso} with pressure information.
\definecolor{colArmFilt}{HTML}{a6cee3}
\definecolor{colArmDes}{HTML}{1f78b4}
\definecolor{colBoomFilt}{HTML}{b2df8a}
\definecolor{colBoomDes}{HTML}{33a02c}
\definecolor{colBucketFilt}{HTML}{fb9a99}
\definecolor{colBucketDes}{HTML}{e31a1c}
\colorlet{colArmLight}{colArmDes!25}
\colorlet{colBoomLight}{colBoomDes!25}
\colorlet{colBucketLight}{colBucketDes!25}
\begin{figure}[tb]
    \centering
    \includegraphics{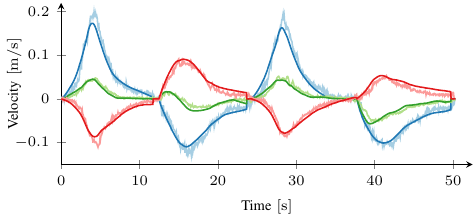}
    \vspace{-1em}
    \caption{Desired (dark) and measured (light) velocity of the arm {\color{colArmDes}(blue)}, boom {\color{colBoomDes}(green)}, and bucket {\color{colBucketDes}(red)} cylinders over the course of one evaluation cycle using the controller derived from the \gls{siso} model~\ref{itm:siso} including pressure information.}
    \label{fig:armVelPlot}
\end{figure}

\section{CONCLUSIONS AND FUTURE WORKS}

We studied the extension of feedforward controller design by feedback linearization of \glspl{lmn} to \gls{mimo} models and disturbance compensation. Further, we derived a criterion for \gls{bibo} stability of the resulting controller for the case where the relative degree is one, which allows the approach to be applied to models with \gls{zd}.
We showed that this method can be used in a complex real world application with small datasets, noisy measurements, and limited computing resources.
Using data collected from a real hydraulic excavator, we trained \glspl{lmn} in a \gls{narx} structure and evaluated the controllers in trajectory tracking hardware experiments. 
The results highlight the effectiveness of our theoretical contributions: The extensions to disturbance compensation, \gls{mimo} systems, and models with \gls{zd} all improve tracking performance compared to the baseline controller from \cite{rolle_2016}.
Further, the advantages of our data-driven approach -- no tuning and few hyperparameters during model training -- are illustrated by our example in a high-mix, low-volume production setting.

While the focus of this work was the \gls{bibo} stability of the feedforward controller, criteria for stronger notions of stability, such as input-to-state stability, can be useful. 
Also, a stability criterion for the controller derived from the \gls{mimo} system could be the focus of future work. 
To ensure stability of the whole system, the dynamics of the outer closed loop have to be considered as well.

\bibliographystyle{IEEEtran}
\bibliography{IEEEabrv,references}

\appendices
\section{Validity Transformation}\label{apx:lemmaForProofStability}
\begin{lemma}\label{lma:activationTransformation}
    Given are the validity vector 
    $
        {\vtActVecA = [
            \vtActA{1} \hdots \vtActA{\lmnNumModels}
        ]^\top}
        \text{ with }
        {\textstyle\sum_{\lmnModelIdx=1}^{\lmnNumModels}\vtActA{\lmnModelIdx}=1
        ,\;}
        {0 \le \vtActA{\lmnModelIdx} \le 1}
    $
    and the parameters
        ${\vtParamVec{\lmnModelIdx} = [
            \vtParam{\narxMaxInpDel}{\lmnModelIdx} \hdots \vtParam{2}{\lmnModelIdx}
        ]^\top,\;}
        {\lmnModelIdx \in [\lmnNumModels]}.$    
    Let all $\vtParam{1}{\lmnModelIdx}\not=0$ for ${\lmnModelIdx \in [\lmnNumModels]}$ have the same sign.
    Then
            $\forall \vtActVecA \exists \vtActVecB:
            \frac{1}{\sum_{\lmnModelIdx=1}^{\lmnNumModels}\vtActA{\lmnModelIdx}\vtParam{1}{\lmnModelIdx}}\left(\textstyle\sum_{\lmnModelIdx=1}^{\lmnNumModels}\vtActA{\lmnModelIdx}\vtParamVec{\lmnModelIdx}\right)
            =
            \textstyle\sum_{\lmnModelIdx=1}^{\lmnNumModels}\frac{\vtActB{\lmnModelIdx}}{\vtParam{1}{\lmnModelIdx}}\vtParamVec{\lmnModelIdx}$
    with $\sum_{\lmnModelIdx=1}^{\lmnNumModels}\vtActB{\lmnModelIdx}=1,\; 0 \le \vtActB{\lmnModelIdx} \le 1$.
\end{lemma}
\vspace{0.5em}
\begin{proof}
    Choose
        $\vtActB{\lmnModelIdx} = \frac{\vtParam{1}{\lmnModelIdx}\vtActA{\lmnModelIdx}}{\sum_{\genIdx=1}^{\lmnNumModels}\vtParam{1}{\genIdx}\vtActA{\genIdx}},$
    such that
    \begin{equation}
            \textstyle\sum_{\lmnModelIdx=1}^{\lmnNumModels}\frac{\vtActB{\lmnModelIdx}}{\vtParam{1}{\lmnModelIdx}}\vtParamVec{\lmnModelIdx} 
            = \frac{1}{\sum_{\genIdx=1}^{\lmnNumModels}\vtParam{1}{\genIdx}\vtActA{\genIdx}}\left(\textstyle\sum_{\lmnModelIdx=1}^{\lmnNumModels}\vtActA{\lmnModelIdx}\vtParamVec{\lmnModelIdx}\right).
    \end{equation}
    With the same sign of all $\vtParam{1}{\lmnModelIdx}$,
        ${\sum_{\lmnModelIdx=1}^{\lmnNumModels}\vtActB{\lmnModelIdx}=1}
        \text{ and }
        {0 \le \vtActB{\lmnModelIdx} \le 1}$
    follows.
\end{proof}

\section{Proof of Open-Loop Stability Criterion}\label{apx:proofStability}
\begin{proof}
    By applying Lemma~\ref{lma:activationTransformation} from Appendix~\ref{apx:lemmaForProofStability} to the last row of the combined system matrix~$\invLmnCombSysMat$ of the controller~\eqref{eq:invLmnStateSpace}, it can be rewritten as
    \begin{equation}
        {\invLmnCombSysMat[\invLmnActAltVec] = \textstyle\sum_{\lmnModelIdx=1}^{\lmnNumModels}\invLmnActAlt{\lmnModelIdx}\invLmnModelFacMat{\lmnModelIdx}\invLmnModelSysMat{\lmnModelIdx}}
    \end{equation}
    with the alternative activation~$\invLmnActAltVec$.
    In~\cite[Thm.~4.2]{tanaka_1992} it has been shown that a fuzzy system with $\lmnNumModels$ linear models 
    \begin{equation}
        {\tsfmStateVec[\timestep+1] = \textstyle\sum_{\lmnModelIdx=1}^{\lmnNumModels} \tsfmAct{\lmnModelIdx} \tsfmModelSysMat{\lmnModelIdx} \tsfmStateVec}
    \end{equation}
    with ${\tsfmStateVec \in \mathbb{R}^{\tsfmNumStates},\;\tsfmModelSysMat{\lmnModelIdx} \in \mathbb{R}^{\tsfmNumStates \times \tsfmNumStates}}$
    and normalized weights 
        ${\sum_{\lmnModelIdx=1}^{\lmnNumModels} \tsfmAct{\lmnModelIdx} = 1
        ,\;}
        {0 \le \tsfmAct{\lmnModelIdx} \le 1}$
    is globally asymptotically stable if 
        ${\exists \posDefMat \succ \bm{0} 
        \quad \tsfmModelSysMat{\lmnModelIdx}^\top \posDefMat \tsfmModelSysMat{\lmnModelIdx} - \posDefMat \prec \bm{0} \; }
        {\forall \lmnModelIdx \in [\lmnNumModels].}$
    By choosing ${\tsfmAct{\lmnModelIdx}=\invLmnActAlt{\lmnModelIdx}}$ and $\tsfmModelSysMat{\lmnModelIdx}=\invLmnModelFacMat{\lmnModelIdx}\invLmnModelSysMat{\lmnModelIdx}$, the claim follows.
\end{proof}

\section{Proof of BIBO Stability of Controller}\label{apx:proofBiboStability}
\begin{proof}
    When applying Lemma~\ref{lma:activationTransformation} to the last rows of $\invLmnCombInpMat$ and $\invLmnCombOffVec$ of the combined affine input and offset 
    \begin{equation}
        \invLmnCombInp = \invLmnCombInpMat \invLmnDesShiftVec + \invLmnCombOffVec,
    \end{equation}
    it can be seen that they can also be represented as sum of constant matrices weighted with alternative validities and thus are bounded.
    Thus, $\invLmnCombInp$ is bounded for any bounded input $\invLmnDesShiftVec$.

    The shift $\invLmnStateVecShift$ of the coordinate transformation of the state ${\invLmnStateVecShifted=\invLmnStateVec+\invLmnStateVecShift}$ can be chosen as 
    \begin{equation}\label{eq:invLmnStateVecShiftChoice}
        \invLmnStateVecShift = \invLmnCombSysMat\invLmnStateVecShift -\invLmnCombInp,
    \end{equation}
    such that 
    \begin{equation}
        \begin{aligned}
            \invLmnStateVec[\timestep+1] &= \invLmnCombSysMat\invLmnStateVec + \invLmnCombInp \\
            \Leftrightarrow\invLmnStateVecShifted[\timestep+1] - \invLmnStateVecShift &= \invLmnCombSysMat(\invLmnStateVecShifted-\invLmnStateVecShift) \\
            &\phantom{={}}+ \invLmnCombInp \\
            \Leftrightarrow\invLmnStateVecShifted[\timestep+1] &= \invLmnCombSysMat\invLmnStateVecShifted
        \end{aligned}
    \end{equation}
    is \gls{gas} with Lemma~\ref{lma:stability}.
    The matrix $\invLmnCombSysMat-\bm{I}$ is non-singular, since the eigenvalues are given as $\lambda-1$, with $\lambda$ as the eigenvalues of $\invLmnCombSysMat$. Due to the \gls{gas}, we know $\lambda \not= 1$ and with~\eqref{eq:invLmnStateVecShiftChoice} follows 
    \begin{equation}
        \invLmnStateVecShift = (\invLmnCombSysMat-\bm{I})^{-1}\invLmnCombInp.
    \end{equation}
    With the matrix $\invLmnCombSysMat-\bm{I}$ as non-singular, its multiplication to a vector provides an invertible and bounded transformation from one Banach space to another. With the bounded inverse theorem follows that its inverse $(\invLmnCombSysMat-\bm{I})^{-1}$ is also bounded.
    Therefore, $\invLmnStateVecShift$ and thus $\invLmnStateVec$ is bounded if $\invLmnDesShiftVec$ and thus $\invLmnCombInp$ is bounded. The output $\dlmnInp$ is part of the state $\invLmnStateVec$.
\end{proof}
\end{document}